\documentclass[10pt,pra,twocolumn,superscriptaddress,showpacs,floatfix]{revtex4}

\usepackage{graphicx}
\usepackage{amssymb}
\usepackage{times}
\usepackage{amsmath}
\usepackage{amsthm}

\newcommand{\bra}[1]{\langle#1|}
\newcommand{\ket}[1]{|#1\rangle}

\begin{document}

\title{Quantum Cloning of Continuous Variable Entangled States}

\author{Christian Weedbrook}\email{christian.weedbrook@gmail.com} \affiliation{Department of Physics, University
of Queensland, St Lucia, Queensland 4072, Australia}

\author{Nicolai B. Grosse} \affiliation{Department of Physics, Australian
National University, Canberra, ACT 0200, Australia}

\author{Thomas Symul} \affiliation{Department of Physics, Australian
National University, Canberra, ACT 0200, Australia}

\author{Ping Koy Lam} \affiliation{Department of Physics, Australian
National University, Canberra, ACT 0200, Australia}

\author{Timothy C. Ralph} \affiliation{Department of Physics, University
of Queensland, St Lucia, Queensland 4072, Australia}

\date{\today}

\begin{abstract}
We consider the quantum cloning of continuous variable entangled states. This is achieved by introducing two symmetric entanglement cloning machines (or e-cloners): a local e-cloner and a global e-cloner; where we look at the preservation of entanglement in the clones under the condition that the fidelity of the clones is maximized. These cloning machines are implemented using simple linear optical elements such as beam splitters and homodyne detection along with squeeze gates. We show that the global e-cloner out-performs the local e-cloner both in terms of the fidelity of the cloned states as well as the strength of the entanglement of the clones. There is a minimum strength of entanglement ($3$dB for the inseparability criterion and $5.7$dB for the EPR paradox criterion) of the input state of the global e-cloner that is required to preserve the entanglement in the clones.
\end{abstract}


\maketitle

\section{Introduction}

A fundamental concept in quantum information theory \cite{Nie00} is the no-cloning theorem \cite{WZ82,Die84}; which ensures that there exists no device capable of perfectly copying an unknown quantum state. The topic of \textit{quantum cloning} is with concerned with finding the best approximation to a quantum copier \cite{Buz96}. These approximate copies or clones are created using a physical apparatus known as a quantum cloning machine. The first of these machines was developed using quantum discrete variables or qubits \cite{Sca05}. Later on this was extended to finite-dimensional systems or qudits \cite{Buz98,Wer98}.

In 2000 the natural extension to the cloning of quantum continuous variables was considered \cite{Cer00}. Continuous variables (CV) \cite{Bra04} commonly deal with the creation, manipulation and processing of Gaussian states \cite{Fer05}. Examples of these include coherent states, squeezed states, thermal states and the Einstein-Podolsky-Rosen (EPR) type of entangled states. They are typically studied due to the ease with which they can be generated and manipulated experimentally, and theoretically analyzed. The first optical implementations of CV cloning machines using parametric amplifiers and beamsplitters were presented in \cite{Bra01,Fiu01,DAr01}. Since then, CV quantum cloning has attracted more and more interest \cite{Cer00a,Coc04,Cer05,Oli06,Sab07}. In \cite{And05} Andersen et al. provided the first experimental demonstration of the cloning of coherent states whilst presenting an even simpler optical implementation using only beamsplitters and homodyne detection. To date, no one has considered the cloning of CV entangled states.

In this paper, we consider the quantum cloning of a class of CV entangled states by analyzing how well the clones have been preserved in terms of their fidelity as well as the strength of their entanglement. We require the entangled states to be copied with the best possible fidelity whilst analyzing how well the entanglement has been preserved in such a situation. We do this by introducing CV quantum entanglement cloning machines, known as \textit{e-cloners}. We consider two different cases: a local e-cloner and a global e-cloner. The local e-cloner consists of two independent cloning machines that copies each arm of the entangled state separately. Whilst the global e-cloner is a single cloning machine that takes the entire entangled state as input and outputs two clones. We show that the global e-cloner can copy the entangled states with a fixed fidelity of $\mathcal{F_G}=4/9$ whilst the local cloner has a fidelity varying as a function of the input squeezing used to make the original entangled states. Furthermore we show that the global e-cloner preserves the strength of the entanglement whilst the local e-cloner never preserves any entanglement in the clones. Finally we give an implementation of these e-cloners using linear optical elements such as beamsplitters, homodyne detection and squeeze gates.

The cloning of entangled states and its entanglement for discrete variables was considered in 1998 by Buzek and Hillery \cite{Buz98,Buz98a} in the form of an arbitrary entangled Bell state. In 2004, Lamoureux et al. \cite{Lam04}, investigated the cloning of an arbitrary unknown maximally entangled state with an extension to the finite-dimensional case given in \cite{Kar05}. Why clone entanglement and entangled states? CV quantum entanglement \cite{Cer07} is used as a central resource in CV quantum information processes such as quantum teleportation \cite{Fur98}, cluster state quantum computation \cite{Men06}, quantum secret sharing \cite{Lan04}, quantum cryptography protocols \cite{Sil02} and eavesdropping attacks \cite{Wee06}. Hence, given its importance in quantum information, the topic of cloning such a resource seems reasonable. Furthermore it is possible that a potential quantum error correction procedure for CV cluster state computation \cite{Men06,van06} might require cloning parts of the cluster. In that case we would be interested in optimally copying both the nodes of the cluster and the entanglement between them to ensure minimal loss of computational power.

This paper is structured as follows. Section II will introduce background concepts such as CV entanglement generation and characterization along with the universal CV linear cloning machine. Section III will reveal the quantum cloning machines for entanglement that we will consider in this paper. The cloning of entanglement will be considered in Section IV with Section V showing how well the entanglement has been preserved in both e-cloners. Section VI will derive the associated fidelities for both e-cloners, with Section VII concluding.

\section{Background}

In this section we introduce background material such as the linear quantum cloning machine for CV and characterize the strength of CV quantum entanglement using two criteria. First though we define the CV quantum information nomenclature used throughout this paper.

\subsection{Notation}

In this paper we use the commutator relation $[\hat{X}^+,\hat{X}^-]= [\hat{x},\hat{p}]=2i$ (i.e. with $\hbar=2$). Hence we can define the quadrature operators of the light field in terms of the creation and annihilation operators as follows
\begin{align}
\hat{X}^{+} &= \hat{x} = \hat{a} + \hat{a}^{\dagger}\\
\hat{X}^{-} &= \hat{p} = i(\hat{a}^{\dagger} - \hat{a})
\end{align}
where ($+$) defines the in-phase or amplitude quadrature and ($-$) the out-of-phase or phase quadrature. The quadrature variance is given by $V^{\pm} = \langle (\hat{X}^{\pm})^2 \rangle - \langle \hat{X}^{\pm} \rangle$. For a coherent state $V^{\pm} = 1$.

\subsection{Continuous Variable EPR Entanglement}

An entangled state in CV is known as an \textit{EPR state} after the famous 1935 Einstein-Podolsky-Rosen paradox paper \cite{Ein35} where entanglement was first discussed. Entanglement can exist in various types of CV systems such as optical modes of light and atomic ensembles (e.g. see \cite{Cer07}). The CV entanglement we will consider cloning in this paper is an \textit{unknown} pure bi-partite Gaussian entanglement \cite{Ade07} using optical modes of light. Here unknown means that this type of EPR state is randomly displaced in phase space. This type of entanglement can be described by the annihilation operators $\hat{a}$ in the following way
\begin{align}
\hat{a}_1 &= \frac{1}{2} (\hat{X}_{epr1}^+ + \hat{X}_{epr1}^-) + \alpha_1\\
\hat{a}_2 &= \frac{1}{2} (\hat{X}_{epr2}^+ + \hat{X}_{epr2}^-) + \alpha_1
\end{align}
where $\alpha$ is a random displacement in phase space and the quadrature amplitudes of each arm of the entanglement are defined as
\begin{align}\label{eq: EPR state1}
\hat{X}^{\pm}_{epr1} &= \frac{1}{\sqrt2} (\hat{X}^{\pm}_{sqz1} +
\hat{X}^{\pm}_{sqz2})\\\label{eq: EPR state2}
\hat{X}^{\pm}_{epr2} &= \frac{1}{\sqrt2} (\hat{X}^{\pm}_{sqz1} -
\hat{X}^{\pm}_{sqz2})
\end{align}
where $\hat{X}^{\pm}_{sqz1}$ and $\hat{X}^{\pm}_{sqz2}$ are two squeezed beams. This type of entanglement can be created from a non-degenerate optical parametric amplification, second-harmonic generation or by interfering two squeezed beams on a $50/50$ beamsplitter.

The beamsplitter approach is what we will consider in this paper and consists of both squeezed beams being squeezed in the amplitude quadrature with one being rotated in phase space by $\pi/2$. These beams are then interfered on a $50/50$ beamsplitter to give the CV entanglement described by Eqs.~(\ref{eq: EPR state1}) and (\ref{eq: EPR state2}) with variances given by
\begin{align}\label{eq: EPR state variance}
V_{epr1}^{\pm} =V_{epr2}^{\pm} =\frac{1}{2} (V_{sqz1}^{\pm} +
V_{sqz2}^{\pm})
\end{align}
Using the following substitutions based on the assumption we have pure states: $V_S \equiv V_{sqz1}^{+} = V_{sqz2}^{-} = V_S^+$ and $1/V_S \equiv V_{sqz1}^{-} = V_{sqz2}^{+} = V_S^-$, we can write the above variance as
\begin{align}
V_{epr} = \frac{1}{2} (V_S + 1/V_S)
\end{align}
where $V_S \in (0,1]$. One characteristic of this type of entanglement is the ease with which it can be disentangled: simply put each arm of the entangled state through a $50/50$ beamsplitter. This disentangling property motivates this class of entanglement because, as we will see, this is an essential feature of the global e-cloner.

\subsubsection{CV Entanglement Criteria}

To classify the strength of the entanglement of the cloned EPR states we will use two common CV entanglement criteria: the inseparability criterion and the EPR paradox criterion. Both criteria rely on the correlation matrix to calculate the strength of the entanglement. A bipartite Gaussian entangled state (and in fact any Gaussian state) can be completely described by its amplitude and phase quadrature coherent amplitudes and its correlation matrix. The correlation matrix contains the first and second order moments of the quadrature operators. It is given by a $4\times4$ matrix \cite{Wal94} where the coefficients of the correlation matrix are given by
\begin{align}\label{eq: CM}
C^{kl}_{mn} = \frac{1}{2} \langle\hat{X}^k_m \hat{X}^l_n +
\hat{X}^l_n \hat{X}^k_m \rangle - \langle \hat{X}^k_m \rangle
 \langle\hat{X}^l_n \rangle
\end{align}
where $\{k,l\}\in \{+,-\}$ and $\{m,n\} \in \{x,y\}$.

The inseparability criterion \cite{Dua00} was developed in 2000 by Duan et al. and involves using elements of the correlation matrix to tell whether two quantum states are entangled (inseparable). In our case, the product form of the inseparability criterion is given by
\begin{align}\label{eq: insep criteria}
\mathcal{I} = \frac{1}{2}\sqrt{C^+_I C^-_I}
\end{align}
where the measurable correlations are defined as
\begin{align}
C^{\pm}_I = C^{\pm \pm}_{xx} + C^{\pm \pm}_{yy} - 2 |C^{\pm \pm}_{xy}|
\end{align}
According to the inseparability criterion, entanglement exists between the two modes $x$ and $y$ when $\mathcal{I}<1$. So for example, a pure EPR state would have $\mathcal{I} = V_S$ where $V_S$ is the variance of the squeezing used to create the entangled state. Hence in that case, a pure EPR state is entangled for all values of squeezing except when the input is a coherent state, i.e. $V_S = 1$.

Another way of classifying the strength of CV entanglement is via the EPR paradox criterion which was introduced in 1988 by Reid and Drummond \cite{Rei88}. The degree of EPR paradox $\epsilon$ which measures the apparent level of violation of the Heisenberg uncertainty principle can also be determined from the coefficients of the correlation matrix. It is defined as
\begin{align}\label{eq: EPR criteria}
  \epsilon = \Big(C^{++}_{xx} - \frac{|C^{++}_{xy}|^2}{C^{++}_{yy}} \Big)
   \Big(C^{--}_{xx} - \frac{|C^{--}_{xy}|^2}{C^{--}_{yy}} \Big)
\end{align}
Here we have EPR violation when the following condition is met, $\epsilon <1$. For example, in the case of a pure input EPR state applying the EPR paradox criteria of Eq.(\ref{eq: EPR criteria}) we find $\epsilon = 4/(V_S + 1/V_S)^2$. So for a coherent state
$\epsilon = 1$ and for any level of squeezing (i.e. $V_S < 1$) we have $\epsilon <1$. It is worth noting in a comparison of the two entanglement criteria, the inseparability criteria picks up a larger class of entangled states and is both necessary and sufficient. On the other hand the EPR paradox criteria is a sufficient but not necessary condition for entanglement between two beams.

Even though the EPR paradox criterion is a stronger requirement we will still consider it as it is commonly used in experiments and is closely related to the original EPR paradox. Additionally, violation of the EPR paradox criterion is a necessary condition to obtain quantum correlations in a CV teleported beam. For example, if the EPR source used in unity gain teleportation does not violate the EPR paradox criterion then it is impossible to get the teleported beam to show quantum mechanical effects, such as squeezing or Wigner function negativity \cite{Gra00a}.

\subsection{Universal Cloning Machine for Continuous Variables}

%
%%%
\begin{figure}[!ht]
\begin{center}
\includegraphics[width=8cm]{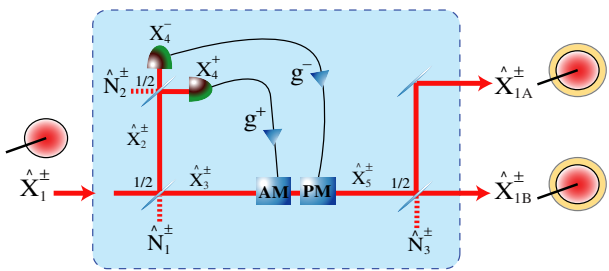}
\caption{A universal continuous variable cloning machine. This symmetric $1 \rightarrow 2$ linear quantum cloning machine perfectly copies the amplitude of all input states (by using unity gain $g^{\pm} = \sqrt2$) whilst adding one unit of vacuum noise as penalty. $\hat{X}^{\pm}_{1}$: input state to be cloned with the cloned output states denoted by: $\hat{X}^{\pm}_{1A}$ and $\hat{X}^{\pm}_{1B}$; AM: amplitude modulator; PM: phase modulator; $\hat{N}^{\pm}_1$, $\hat{N}^{\pm}_2$ and $\hat{N}^{\pm}_3$: vacuum noise terms.} \label{figure_linear_cloner}
\end{center}
\end{figure}

The quantum cloning machine for CV is known as a \textit{linear cloning machine} and was introduced in \cite{And05} and will form a central role in our entanglement cloning machines later on. It consists of simple linear optical elements such as beamsplitters and homodyne detection with feed forward (c.f. Fig.\ref{figure_linear_cloner}). The gain of the feed forward is commonly set such as to achieve unit gain for the entire circuit, where in our case unity gain is achieved by setting $g^{\pm} = \sqrt2$. In this way the amplitude of any state is copied perfectly but with a unit of noise added as a penalty. Hence unity gain allows us to maximize the fidelity of the output stated compared to the input state. It is in this sense that the linear cloning machine is a universal cloner. In Fig.\ref{figure_linear_cloner}, the final $50/50$ beamsplitter is used to reduce the amplitude to unity and to output the two clones ($A$ and $B$) of the original state $\hat{X}^{\pm}_1$ which results in $\hat{X}_{1A}^{\pm} = \hat{X}_1^{\pm} - \hat{N}^{\pm}_4$ and $\hat{X}_{1B}^{\pm} = \hat{X}_1^{\pm} + \hat{N}^{\pm}_5$ where $\hat{N}^{\pm}_4 $ and $\hat{N}^{\pm}_5$ are overall combination of noise terms from the cloning circuit. We observe firstly that, because the noise terms have zero mean, the first order moments of the cloner are identical to the original, i.e. $\langle \hat{X}_{1A}^{\pm }\rangle = \langle \hat{X}_{1B}^{\pm }\rangle = \langle \hat{X}_{1}^{\pm }\rangle$. On the other hand, for the second order moments we obtain
\begin{align}
V_{1A}^{\pm} = V_{1B}^{\pm} = V_1^{\pm} + 1
\end{align}
where we assume the cross terms are not correlated and do not contribute and the variance of the noise terms are set to one. Hence the universal CV cloning machine adds one unit of quantum noise to the output cloned states whilst perfectly copying the classical amplitude. This linear cloning machine has been used to optimally clone coherent states and squeezed states (with a known squeezing parameter) along with other Gaussian states \cite{Oli06}.

\section{Quantum Cloning Machines for Continuous Variable Entanglement}

\begin{figure}[!ht]
\begin{center}
\includegraphics[width=8cm]{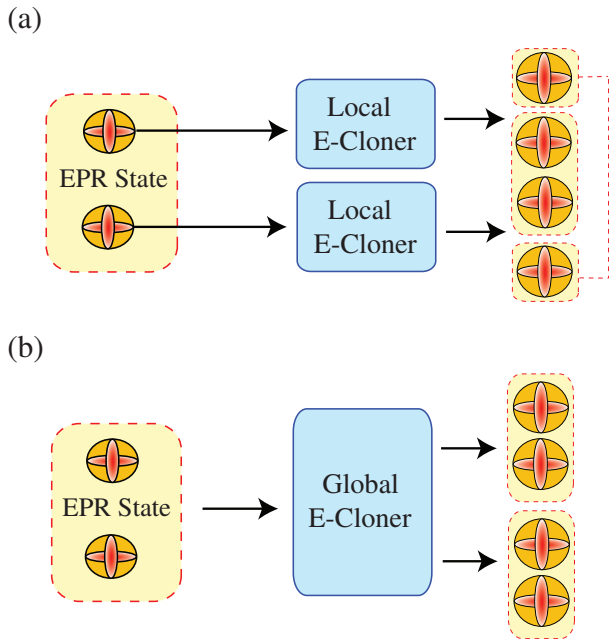}
\caption{Continuous variable entanglement cloning machines (e-cloners) used to clone the entangled states as well as the entanglement itself. (a) Local e-cloner consists of two individual cloning machines one for each arm of the entanglement. (b) The global e-cloner takes the whole entangled state and outputs the imperfect cloned copies of the initial entangled state.}\label{figure_ecloners}
\end{center}
\end{figure}

In this section we introduce two CV entanglement cloning machines: (1) the local e-cloner and (2) the global e-cloner. These machines are symmetric $1 \rightarrow 2$ quantum cloning machines, i.e. the fidelities are identical for both clones and we have two copies of our initial input state. In both cases, the input state is chosen to be a randomly displaced EPR state in order to make the input state \textit{unknown}. The unknown displacements are random shifts in phase space in both the position and momentum quadratures. Figure~\ref{figure_ecloners} gives a graphic of the two cloning machines. The local e-cloner Fig.~\ref{figure_ecloners}(a) consists of two local cloning machines that clone each arm of the entanglement separately. On the other hand the global e-cloner Fig.~\ref{figure_ecloners}(b) is a single cloning machine that takes the entire entanged state as input, and outputs two copies. We will now introduce each e-cloner in more detail via an optical implementation which has been chosen based on the linear cloning machine introduced in Section II and requires a simple setup with minimal resources. Although, we point out, that the quantum cloning machines we introduce in this paper and the associated unitary operations of the quantum cloning circuits, can be directly applied to any CV system, e.g. the quantum cloning of entangled atomic ensembles.

\subsection{Local E-Cloner}

\begin{figure}[!ht]
\begin{center}
\includegraphics[width=8cm]{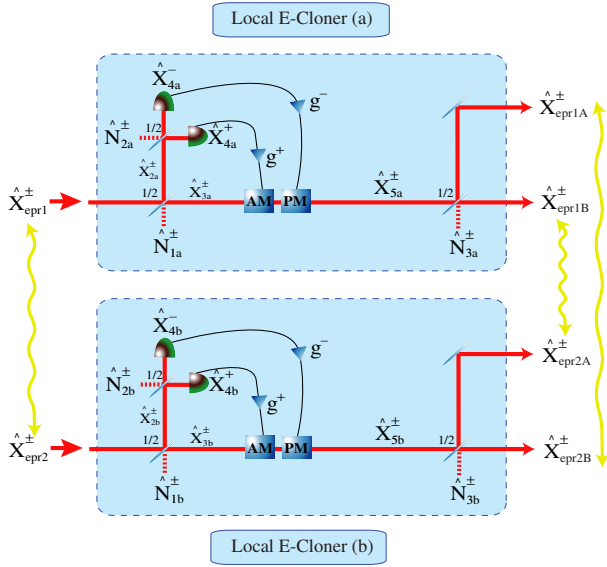}
\caption{Schematic of the local e-cloner. Each arm of the entangled state is cloned separately using a linear cloning machine which copies the amplitude perfectly whilst adding one unit of vacuum noise onto each clone. The local e-cloner is a universal continuous variable cloning machine. The yellow lines indicates which arms are entangled. See the schematic of Fig.~\ref{figure_linear_cloner} for details on the optical elements.}\label{figure_local_ecloner}
\end{center}
\end{figure}

Figure~\ref{figure_local_ecloner} gives a schematic of the local entanglement cloning machine. The local e-cloner must be universal because it consists of two universal linear cloning machines as described in Section II and featured in Fig.~\ref{figure_linear_cloner}. Here a randomly displaced EPR state is created with one arm of the entanglement sent through a local e-cloner and the other arm sent through the other e-cloner. This setup corresponds to cloning two thermal states and adds one unit of vacuum noise to each of the original input states. Note that if the initial channel transmission was $\eta=0$ instead of $\eta=1/2$ then we just get the $50/50$ beamsplitter cloning machine. This is a kin to a ``classical approach" to quantum cloning where we measure as best we can both quadratures simultaneously and then displace a newly created EPR state according to the measurement results. However the problem with this approach is that we introduce two units of vacuum noise to our cloning machine which does not give us the best possible fidelity.

We also point out that for both e-cloners we require that the fidelity of the clones is optimized. Specifically, this corresponds to having the gain set to unity because if we take our input set from a large ensemble, then the fidelity will reduce significantly if we do not operate at unity gain \cite{Bra00,Gra00a}. Also if we did not consider the fidelity of the clones, then we could always build a ``cloning machine" that re-created new entanglement (independent of the input) with as much entanglement as physically possible.

\subsection{Global E-Cloner}

The fundamental difference between the synthesizing of the local e-cloner compared to the global e-cloner is the requirement of six additional in-line squeezers or squeeze gates for the global cloning machine. The squeezed gates is the reason why the global e-cloner is not a universal cloning machine as it contains a variable parameter, i.e. we need to know how entangled the input state is in order to know how much to un-squeeze the input states. Therefore we have to vary the operation of the squeezed gates which therefore makes it input state dependent or non-universal. This is in contrast to the universal local e-cloner where all parts of the cloning machine are fixed.

\begin{figure}[!ht]
\begin{center}
\includegraphics[width=8.5cm]{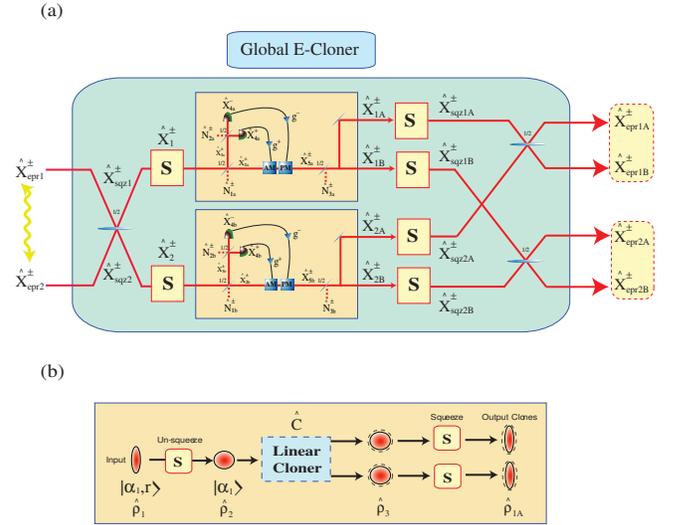}
\caption{Schematic of the global e-cloner. The whole entangled state is cloned producing two imperfect cloned outputs. One of the main differences between the global and local e-cloner is the addition of the squeeze gates along with the fact that the global e-cloner is not a universal cloning machine but rather a state dependant cloner. (a) Schematic of the global e-cloner where $S$ is a squeeze gate. The global e-cloner is made up of two linear cloning machines. See text for details on the optical schematics. (b) A pictorial view of what is happening to each arm of the entangled input state of the global e-cloner in terms of cloning a squeezed state with known squeezing parameter.}\label{figure_global_cloner}
\end{center}
\end{figure}

Squeeze gates form an important component in CV quantum computation \cite{Llo99,Men06,van06} and have recently been conditionally \cite{Jeo06} and deterministically \cite{Yos07} created. The global e-cloner is represented by Fig.~\ref{figure_global_cloner}(a). Here the state to be cloned enters the machine and is immediately disentangled into the two original squeezed states that were used to create the entanglement. These squeezed states are then both ``un-squeezed" into coherent states and cloned using the linear cloning machine. If the strength of the squeezing, given by the squeezing parameter, is known (as it is in our case) then this un-squeezing of the squeezed state into a coherent state followed by cloning the resulting coherent state, is the optimal (Gaussian) strategy \cite{Cer00,Oli06} (c.f. Fig.~\ref{figure_global_cloner}(b)). Finally the two sets of two clones are then squeezed again using the same amount of squeezing that we used to un-squeeze them initially.

It is important that we do not use more squeezing than what was initially used to un-squeeze, otherwise the quantum cloning machine would in a sense be ``cheating" by adding more entanglement than what was there initially. We then have four squeezed states with one unit of noise added on each of them. Finally we then interfere the appropriate squeezed state clones on a $50/50$ beamsplitter to create two entangled states. These are then the clones of the initial entangled state.


\section{Cloning Continuous Variable Entanglement}

In this section we derive the output cloned states of both the local and global e-cloners. To do this we follow the evolution of the quantum operators through the e-cloners in the Heisenberg picture. We can then describe the final output in the Heisenberg picture which will be used in the following section to calculate the strength of the entanglement using the inseparability and EPR paradox criteria. Again we point out that the input states are initially randomly displaced in phase space by two variables $\mathcal{S}^+$ and $\mathcal{S}^-$ chosen from a Gaussian distribution with zero mean. However in the following calculations we can neglect these classical terms as they will not affect the strength of the entanglement.

\subsection{Local E-Cloner}

The two input states to our cloning machine are the entangled states described by Eqs.~(\ref{eq: EPR state1}) and (\ref{eq: EPR state2}) with corresponding variances given by Eq.~(\ref{eq: EPR state variance}). Lets look at the cloning of the first arm $\hat{X}^{\pm}_{epr1}$ initially. Following the notation given in the schematic of Fig.~\ref{figure_local_ecloner}, we see that the two
states after the first $50/50$ beamsplitter are given by
\begin{eqnarray}\label{Operator_After_First_BS}
\hat{X}_{2a}^{\pm} = \frac{1}{\sqrt2}(\hat{X}^{\pm}_{epr1} - \hat{N}_{1a}^{\pm})\\
\hat{X}_{3a}^{\pm} = \frac{1}{\sqrt2}(\hat{X}^{\pm}_{epr1} +
\hat{N}_{1a}^{\pm})
\end{eqnarray}
Hence we can write the equations describing the quantum states
after the second $50/50$ beamsplitter as
\begin{eqnarray}\label{Operator_After_Second_BS}
\hat{X}_{4a}^{\pm} = \frac{1}{\sqrt2} \Big(\frac{1}{\sqrt2}
(\hat{X}^{\pm}_{epr1} - \hat{N}_{1a}^{\pm})\mp
\hat{N}_{2a}^{\pm})\Big)
\end{eqnarray}
The state $\hat{X}_{5a}^{\pm} = \hat{X}_{3a}^{\pm} + g^{\pm} \hat{X}_{4a}^{\pm}$ and is
given by
\begin{eqnarray}\label{Operator_After_Gain}
\hat{X}_{5a}^{\pm} = \frac{1}{\sqrt2}(\hat{X}^{\pm}_{epr1} +
\hat{N}_{1a}^{\pm}) + g^{\pm}\hat{X}_{4a}^{\pm}
\end{eqnarray}
where $g^{\pm}$ can be thought of as some experimental gain used to cancel the
noise terms $\hat{N}_1^{\pm}$ from the e-cloner. We now need to optimize the gain
$g^{\pm}$. To do this we use Eq.~(\ref{Operator_After_Gain}) with
Eq.~(\ref{Operator_After_Second_BS}) to give
\begin{eqnarray}\nonumber
\hat{X}_{5a}^{\pm} = \frac{1}{\sqrt2}(\hat{X}^{\pm}_{epr1} +
\hat{N}_{1a}^{\pm}) + \frac{g^{\pm}}{\sqrt2}\Big(\frac{1}{\sqrt2}
(\hat{X}^{\pm}_{epr1} - \hat{N}_{1a}^{\pm})\mp \hat{N}_{2a}^{\pm})\Big)
\end{eqnarray}
Re-arranging and collecting like terms we get
\begin{eqnarray}\nonumber
\hat{X}_{5a}^{\pm} = \hat{X}^{\pm}_{epr1}
\Big(\frac{g^{\pm}}{2} + \frac{1}{\sqrt2}\Big) +
\hat{N}_{1a}^{\pm} \Big(\frac{1}{\sqrt2} - \frac{g^{\pm}}{2} \Big) \mp
\frac{g^{\pm}}{\sqrt2}\hat{N}_{2a}^{\pm}
\end{eqnarray}
Therefore we can see that to cancel the noise term $N_{1a}^{\pm}$ we need
to let $g^{\pm} = \sqrt2$. This will be our value of unity gain and is chosen as such in order to maximize the fidelity of the output clones. We then have
\begin{eqnarray}
\hat{X}_{5a}^{\pm} = \sqrt2 \hat{X}^{\pm}_{epr1} \mp
\hat{N}_{2a}^{\pm}
\end{eqnarray}
After the last $50/50$ beamsplitter we have the two clones ($A$ and $B$) of $\hat{X}_{epr1}^{\pm}$ described by
\begin{eqnarray}
\hat{X}^{\pm}_{epr1A} = \hat{X}^{\pm}_{epr1} + \hat{N}_{4}^{\pm}\\
\hat{X}^{\pm}_{epr1B} = \hat{X}^{\pm}_{epr1} + \hat{N}_{5}^{\pm}
\end{eqnarray}
where $N^{\pm}_{4} = (\mp \hat{N}_{2a}^{\pm} - \hat{N}_{3a}^{\pm})/ \sqrt2$ and $N^{\pm}_{5} = (\mp \hat{N}_{2a}^{\pm} + \hat{N}_{3a}^{\pm})/ \sqrt2$ are the overall noise terms given by a combination of all the previously introduced noise. Therefore we have (imperfectly) cloned one arm of the initial entangled state. The other two clones of $\hat{X}^{\pm}_{epr2}$ are calculated in the same way and are given by
\begin{align}
\hat{X}^{\pm}_{epr2A} = \hat{X}^{\pm}_{epr2} + \hat{N}^{\pm}_{6}\\
\hat{X}^{\pm}_{epr2B} = \hat{X}^{\pm}_{epr2} + \hat{N}^{\pm}_{7}
\end{align}
where $N^{\pm}_{6} = (\mp \hat{N}_{2b}^{\pm} - \hat{N}_{3b}^{\pm})/ \sqrt2$ and $N^{\pm}_{7} = (\mp \hat{N}_{2b}^{\pm} + \hat{N}_{3b}^{\pm})/ \sqrt2$ are the different (i.e. independent) noise terms from the other local e-cloner. We are now in a position to write the final two sets of operators describing the outputs from the local e-cloner. The first cloned EPR state is (c.f. Fig.~\ref{figure_local_ecloner})
\begin{align}\nonumber
\hat{X}^{\pm}_{epr1A} = \hat{X}^{\pm}_{epr1} + \hat{N}_4^{\pm}\\\label{eq: cloned EPR state local}
\hat{X}^{\pm}_{epr2B} = \hat{X}^{\pm}_{epr2} + \hat{N}^{\pm}_7
\end{align}
and the second cloned EPR state
\begin{align}\nonumber
\hat{X}^{\pm}_{epr1B} = \hat{X}^{\pm}_{epr1} + \hat{N}_5^{\pm}\\\label{eq: cloned EPR state local 2}
\hat{X}^{\pm}_{epr2A} = \hat{X}^{\pm}_{epr2} + \hat{N}^{\pm}_6
\end{align}
Therefore calculating the variances we have
\begin{align}\label{eq: local cloner output}\nonumber
V_{epr 1A(B)}^{\pm} = \langle (\hat{X}_{epr1A}^{\pm})^2
\rangle &=
\langle (\hat{X}_{epr1B}^{\pm})^2 \rangle\\\nonumber
&= V_{epr1}^{\pm} + V_N^{\pm}\\
&= V_{epr1}^{\pm} + 1
\end{align}
which due to symmetry is also equal to the other two clones, i.e. $V_{epr2A(B)}^{\pm} = V_{epr1}^{\pm} + 1$. We also know the cross terms are not correlated and are therefore equal to zero and in the last line we have the symmetry $V^{\pm}_{epr1} = V^{\pm}_{epr2}$. Also the variance of all the noise terms are set to one, i.e. $V(\hat{N}_2^{\pm}) = V(\hat{N}_3^{\pm}) = 1$ which gives $V_N^{\pm} = (V(\hat{N}_2^{\pm}) + V(\hat{N}_3^{\pm}))/2 = 1$. Consequently we can see from the above set of equations that the final two clones for each arm of the EPR state consists of the initial input state with a penalty of one unit of noise from the local entanglement cloning machine.

\subsection{Global E-Cloner}

We will now analyze the global e-cloner in the same way we did for the local e-cloner. The two input states of our cloning machine are again the entangled states described by Eqs.~(\ref{eq: EPR state1}) and (\ref{eq: EPR state2}) with corresponding variances given in Eq.~(\ref{eq: EPR state variance}). As they first enter the global e-cloner, they encounter a $50/50$ beamsplitter whose purpose is to disentangle them into the original two input squeezed states used to create the entanglement, i.e. $\hat{X}^{\pm}_{sqz1}$ and $\hat{X}^{\pm}_{sqz2}$. Our next goal is to optimally clone these two squeezed states. However, as mentioned before, a squeezer followed by an optimal (Gaussian) coherent state cloner and another squeezer is equivalent to an optimal (Gaussian) squeezed state cloner in the case of the squeezing parameter being known. When we first un-squeeze the squeezed states $\hat{X}^{\pm}_{sqz1}$ and $\hat{X}^{\pm}_{sqz2}$ by an amount $s^{\pm}$ [where $s^{\pm} \in (0,1]$ and the variance of $s^{\pm}$ (i.e. $V_S^{\pm}$) is the same variance as the squeezed states used to create the original EPR state (i.e. $V_S^{\pm} = V_{sqz1}^{\pm} = V_{sqz2}^{\mp}$)] we have the resulting coherent states described by
\begin{align}
\hat{X}_{1}^{\pm} &= \frac{1}{s^{\pm}} \hat{X}^{\pm}_{sqz1}\\
\hat{X}_{2}^{\pm} &= s^{\pm} \hat{X}^{\pm}_{sqz2}
\end{align}
Due to the symmetry of the two, we can look at putting $\hat{X}^{\pm}_1$ through the linear cloning machine which is described in Section II. The two clones ($A$ and $B$) of this input state from the linear cloning machine are given by
\begin{align}
\hat{X}_{1A}^{\pm} &= \frac{1}{s^{\pm}} \hat{X}^{\pm}_{sqz1} - \hat{N}_4^{\pm}\\
\hat{X}_{1B}^{\pm} &= \frac{1}{s^{\pm}} \hat{X}^{\pm}_{sqz1} + \hat{N}_5^{\pm}
\end{align}
The next step is to squeeze these cloned coherent states back into squeezed states using the same amount of squeezing that we had used to un-squeeze them, i.e. $s^{\pm} \hat{X}^{\pm}_{1A}$ and $s^{\pm} \hat{X}^{\pm}_{1B}$. Hence we have
\begin{align}
\hat{X}^{\pm}_{sqz1A} &= \hat{X}^{\pm}_{sqz1} - s^{\pm} \hat{N}^{\pm}_4\\
\hat{X}^{\pm}_{sqz1B} &= \hat{X}^{\pm}_{sqz1} + s^{\pm} \hat{N}^{\pm}_5
\end{align}
where $\hat{X}^{\pm}_{sqz1A}$ and $\hat{X}^{\pm}_{sqz1B}$ are the cloned copies of
$\hat{X}^{\pm}_{sqz1}$ and $s^{\pm} \hat{N}^{\pm}_4$ and $s^{\pm} \hat{N}^{\pm}_5 $ are the combined noise terms from the imperfect cloning process which are now squeezed. Following the same procedure, we have the two cloned copies of the other squeezed state $\hat{X}^{\pm}_{sqz2}$ given by
\begin{align}
\hat{X}^{\pm}_{sqz2A} &= \hat{X}^{\pm}_{sqz2} - \frac{1}{s^{\pm}} \hat{N}^{\pm}_6\\
\hat{X}^{\pm}_{sqz2B} &= \hat{X}^{\pm}_{sqz2} + \frac{1}{s^{\pm}} \hat{N}^{\pm}_7
\end{align}
where again $\hat{N}^{\pm}_6/s^{\pm}$ and $\hat{N}^{\pm}_7/s^{\pm} $ are the combined (anti-squeezed) noise terms. Recombining the appropriate clones of the squeezed states back onto a $50/50$ beamsplitter to create entanglement (c.f. Fig.~\ref{figure_global_cloner}(a)) will give us
\begin{align}\nonumber
X_{epr1A}^{\pm} &= \frac{1}{\sqrt2}(X_{sqz1A}^{\pm} + X_{sqz2A}^{\pm})\\\nonumber
X_{epr1B}^{\pm} &= \frac{1}{\sqrt2}(X_{sqz1A}^{\pm} - X_{sqz2A}^{\pm})\\
X_{epr2A}^{\pm} &= \frac{1}{\sqrt2}(X_{sqz1B}^{\pm} + X_{sqz2B}^{\pm})\\\nonumber
X_{epr2B}^{\pm} &= \frac{1}{\sqrt2}(X_{sqz1B}^{\pm} - X_{sqz2B}^{\pm})
\end{align}
Hence the final cloned arms of the original EPR state can be written as
\begin{align}\nonumber
X_{epr1A}^{\pm} &= \frac{1}{\sqrt2}(X_{sqz1}^{\pm} + X_{sqz2}^{\pm}) + (-s^{\pm} N_4^{\pm} - \frac{1}{s^{\pm}} N_6^{\pm})/\sqrt2\\\nonumber
X_{epr1B}^{\pm} &= \frac{1}{\sqrt2}(X_{sqz1}^{\pm} - X_{sqz2}^{\pm}) + (-s^{\pm} N_4^{\pm} + \frac{1}{s^{\pm}} N_6^{\pm})/\sqrt2\\\nonumber
X_{epr2A}^{\pm} &= \frac{1}{\sqrt2}(X_{sqz1}^{\pm} + X_{sqz2}^{\pm}) + (s^{\pm} N_5^{\pm} + \frac{1}{s^{\pm}} N_7^{\pm})/\sqrt2\\
X_{epr2B}^{\pm} &= \frac{1}{\sqrt2}(X_{sqz1}^{\pm} -
X_{sqz2}^{\pm}) + (s^{\pm} N_5^{\pm} - \frac{1}{s^{\pm}} N_7^{\pm})/\sqrt2
\end{align}
We can rewrite the above equations into the more familiar form containing the arms of the original entangled input state, i.e. $\hat{X}^{\pm}_{epr1}$ and $\hat{X}^{\pm}_{epr2}$. Accordingly both arms of the first clone of the original EPR state using the global e-cloner are given by
\begin{align}\nonumber
X_{epr1A}^{\pm} &= \hat{X}_{epr1}^{\pm} + (-s^{\pm} N_4^{\pm} - \frac{1}{s^{\pm}} N_6^{\pm})/\sqrt2\\\label{eq: cloned EPR state global 1}
X_{epr1B}^{\pm} &=  \hat{X}_{epr2}^{\pm} + (-s^{\pm} N_4^{\pm} + \frac{1}{s^{\pm}} N_6^{\pm})/\sqrt2
%%
%%
%X_{epr2A}^{\pm} &=  \hat{X}_{epr1}^{\pm} - (N_1^{\pm} + N_2^{\pm})/\sqrt2\\\nonumber
%%
%%
%X_{epr2B}^{\pm} &=  \hat{X}_{epr2}^{\pm} - (N_1^{\pm} - N_2^{\pm})/\sqrt2
\end{align}
%
%where $\hat{N}_8^{\pm} = (-N_4^{\pm} - N_6^{\pm})/\sqrt2$ and $\hat{N}_9^{\pm} = (-N_4^{\pm} + N_6^{\pm})/\sqrt2$.
The second clone can be described as
\begin{align}\nonumber
%X_{epr1A}^{\pm} &= \hat{X}_{epr1}^{\pm} + (N_1^{\pm} + N_2^{\pm})/\sqrt2\\\nonumber
%%
%%
%X_{epr1B}^{\pm} &=  \hat{X}_{epr2}^{\pm} + (N_1^{\pm} - N_2^{\pm})/\sqrt2\\
%%
%%
X_{epr2A}^{\pm} &=  \hat{X}_{epr1}^{\pm} + (s^{\pm} N_5^{\pm} + \frac{1}{s^{\pm}} N_7^{\pm})/\sqrt2\\\label{eq: cloned EPR state global 2}
X_{epr2B}^{\pm} &=  \hat{X}_{epr2}^{\pm} + (s^{\pm} N_5^{\pm} - \frac{1}{s^{\pm}} N_7^{\pm})/\sqrt2
\end{align}
We can now see one of the differences between the local e-cloner and the global e-cloner. The global e-cloner has noise contributions from both the two internal linear cloning machines. On the other hand the local e-cloner is stuck with the noise terms from the separate linear cloning machines and is unable to swap the noise terms from both. The above equations, when written in terms of their variances, will have the following form
\begin{align}\nonumber
V_{epr1A}^{\pm} &= V_{epr2A}^{\pm} =V_{epr1}^{\pm} + \frac{1}{2} (V_S^{\pm}+ \frac{1}{V_S^{\pm}})\\\label{eq: variance_global_clones}
V_{epr1B}^{\pm} &= V_{epr2B}^{\pm} =  V_{epr2}^{\pm} + \frac{1}{2} (V_S^{\pm}+ \frac{1}{V_S^{\pm}})
\end{align}
In the end all clones from the global e-cloner due to symmetry will have the following form
\begin{align}
 V_{eprA(B)}= V_S + \frac{1}{V_S}
\end{align}
where we have symmetrized both quadratures to give $V_S \equiv V_{sqz1}^{+} = V_{sqz2}^{-} = V_S^+$ and $1/V_S \equiv V_{sqz1}^{-} = V_{sqz2}^{+} = V_S^-$. The above equation shows that the variance of the output of the clones is twice the variance of the input of the original EPR state.

If we compare Eq.(\ref{eq: variance_global_clones}) of the global e-cloner  to the clones from the local e-cloner given in Eq.~(\ref{eq: local cloner output}) we see that the local cloner adds one unit of noise to the original state. On the other hand the global e-cloner adds the average of the squeezed noise term with the anti-squeezed noise term from the squeeze gates of the global e-cloner. We will now use the equations that were derived in this section to determine how well the entanglement has been preserved in the clones from the local and global e-cloners.

\section{Preservation of Entanglement in Clones}

We are now in a position to calculate how well the entanglement has been preserved in the clones from the two e-cloners. The previous section showed that the cloned outputs of the local e-cloner were described by Eqs. (\ref{eq: cloned EPR state local}) and (\ref{eq: cloned EPR state local 2}). However, due to symmetry, we can choose either pair to analyze; likewise for the global cloning machine where Eqs.(\ref{eq: cloned EPR state global 1}) and (\ref{eq: cloned EPR state global 2}) describe its cloned outputs. The correlation matrices (c.f. Eq.~(\ref{eq: CM})) for the outputs of both e-cloners can be calculated and are given by
\begin{widetext}
\begin{align}
CM_L&=
\left(%
\begin{array}{cccc}
  \frac{1}{2} (V_S+1/V_S) +1 & 0 & \frac{1}{2}(V_S - 1/V_S) & 0 \\
  0 & \frac{1}{2} (V_S+1/V_S) +1 & 0 & \frac{1}{2}(1/V_S - V_S) \\
   \frac{1}{2}(V_S - 1/V_S) & 0 &  \frac{1}{2} (V_S+1/V_S) +1 & 0 \\
  0 & \frac{1}{2}(1/V_S - V_S) & 0 & \frac{1}{2} (V_S+1/V_S) +1 \\
\end{array}%
\right)\\\nonumber\\
CM_G&=
\left(%
\begin{array}{cccc}
  V_S+ 1/V_S & 0 & V_S - 1/V_S & 0 \\
  0 & V_S+ 1/V_S & 0 & 1/V_S - V_S \\
  V_S - 1/V_S & 0 & V_S+ 1/V_S & 0 \\
  0 & 1/V_S - V_S & 0 & V_S+ 1/V_S \\
\end{array}%
\right)
\end{align}
\end{widetext}
where we have used the fact that all cross-correlations are zero between any two modes and the individual modes themselves, i.e. $\langle \hat{X}^{\pm}_{sqz1(2)} \hat{X}^{\mp}_{sqz1(2)}\rangle =\langle \hat{X}^{\pm}_{sqz1(2)} \hat{X}^{\pm}_{sqz2(1)}\rangle =0$. This results in $C^{\pm \mp}_{xy} = C^{\pm \mp}_{yx} = 0$ and, due to symmetry, $C^{\pm \pm}_{xy} = C^{\pm \pm}_{yx}$.

\begin{figure}[!ht]
\begin{center}
\includegraphics[width=8cm]{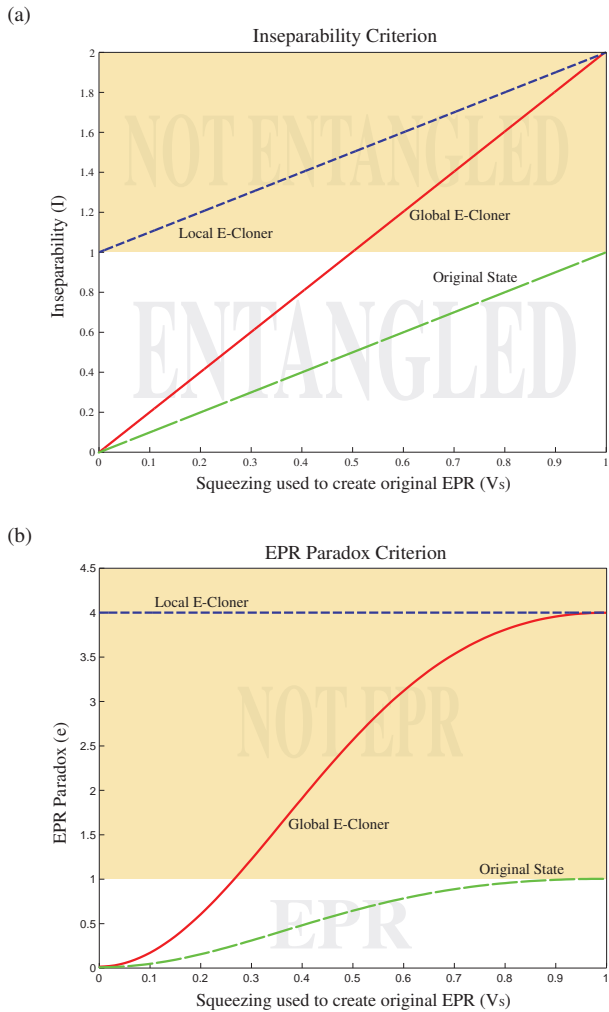}%figure_results
\caption{Plots of the entanglement criteria for both e-cloners. The performance of the local e-cloner (blue dashed line) and the global e-cloner (red solid line) in terms of (a) the inseparability criterion ($\mathcal{I_L}$ and $\mathcal{I_G}$) and (b) the EPR paradox criterion ($\epsilon_L$ and $\epsilon_G$). In both cases the bottom axis is the amount of squeezing used to create the initial entanglement with $V_S = 0$ being perfect squeezing and $V_S = 1$ being a coherent state. The global e-cloner is capable of preserving entanglement according to both criteria whereas the local e-cloner destroys all entanglement in both cases. Here the original entangled input state to be cloned is given for comparison as the green (long dashed) line.}\label{figure_results_1}
\end{center}
\end{figure}

Using the above correlation matrices along with Eqs.(\ref{eq: insep criteria}) and (\ref{eq: EPR criteria}) we can calculate the inseparability criteria ($\mathcal{I_L}$) and EPR paradox criteria ($\epsilon_L$) for the local e-cloner as
\begin{align}
\mathcal{I_L} &= V_S + 1\\
\epsilon_L &= 4
\end{align}
The corresponding equations for the global e-cloner are given by
\begin{align}
\mathcal{I_G} &= 2V_S\\
\epsilon_G &= \frac{16}{(V_S+ 1/V_S)^2}
\end{align}
The above results are plotted in Fig.~\ref{figure_results_1}. In both cases we find that the global e-cloner, out-performs the local e-cloner according to both criteria. In fact, the local e-cloner does not show any preservation of entanglement when using either of the two criteria.

Figure \ref{figure_results_1} (a) shows the performance of the local and global e-cloners in terms of the inseparability criteria. For the global e-cloner, perfect entanglement (i.e. $V_S=0$) is preserved (i.e. $\mathcal{I} =0$). This is because if we start off with an infinitely squeezed source which then is unsqueezed using an infinite amount of energy and then we clone the coherent states using the linear cloner adding one unit of noise. These mixed states are then perfectly squeezed, using an infinite amount of energy, back to perfectly squeezed states which in turn are perfectly entangled states. The line then crosses from the entangled region to the not entangled region when the squeezing used to create the initial entanglement is $V_S=1/2$. In this case if $V_S=1/2$ the global e-cloner unsqueezes it to give a coherent state $V_S=1$. The linear cloning part then adds one unit of noise to both cloners giving $V_S=2$ and then squeezing back again gives $V_S=1$. This corresponds to two coherent states which are not entangled. On the other hand, the local e-cloner destroys all the entanglement during the cloning process so that no matter how much entanglement one starts off with none survives. It initially starts off at $\mathcal{I}=1$ corresponding to the one unit of noise added by the cloner. In the end both, e-cloners asymptote to the same value for the case of a coherent state, because at that stage they are the same cloning machine (the global e-cloner does not need to use the squeeze gates).

Figure \ref{figure_results_1}(b) shows the performance of the local and global e-cloners in terms of the EPR paradox criterion where again the global e-cloner out-performs the local e-cloner. Here the local e-cloner stays at a fixed value of $\epsilon =4$ exhibiting no entanglement irrespective of the strength of the entanglement of the initial EPR state. Like the inseparability criterion plot, both entanglement cloning machines asymptote to the same value for the coherent state input case. This corresponds to four times the value of $\epsilon$ for a pure EPR state which comes from two units of vacuum for each of the conditional variances giving a value of $\epsilon = 4$. The global e-cloner crosses over from the entangled to the separable region (i.e. $\epsilon = 1$) for a value of $V_S = 0.67$ which corresponds to 5.7dB of squeezing.

The global e-cloner perform better than the local e-cloner because the global e-cloner allows the swapping of noise terms between the two local cloning machines inside the global e-cloner which helps preserve the quantum correlations and hence provide stronger entanglement. So essentially the global e-cloner could be thought of as a machine that creates entanglement from two mixed or noisy squeezed states. This is in direct contrast to the local e-cloner which does not allow for this swapping and simply adds one unit noise to each arm of the EPR state thereby destroying the entanglement completely.

\section{Fidelities of Cloned EPR States}

We would now like to determine the quality of the clones produced by these e-cloners. This is achieved by calculating the fidelities of the clones for both the local and global e-cloners. The fidelity $\mathcal{F}$ is a measure of how well the output clones compare to the original input state. Mathematically it is given by the overlap of the pure input state $\ket{\psi}$ with the mixed cloned output state $\hat{\rho}_{out}$
\begin{align}\label{eq: fidelity}
\mathcal{F} = \bra{\psi} \hat{\rho}_{out} \ket{\psi}
\end{align}
We will now calculate the fidelities of the local and global e-cloners.

\subsubsection{Local E-Cloner Fidelity}

The fidelity formula given in Eq.~(\ref{eq: fidelity}) tells us that we need to first determine $\hat{\rho}_{out}$ (the cloned output states) for the local e-cloner. Using the fidelity techniques given in \cite{Cer00,Gra00} we can describe this mixed output state as
\begin{align}\nonumber
\hat{\rho}_{out}
&= \int dx_i dp_i \hspace{1.5mm}G(x_i,p_i)
\hspace{1.5mm} \hat{D}_1 \otimes \hat{D}_2 \ket{\psi} \bra{\psi} \hat{D}_1^{\dagger} \otimes \hat{D}_2^{\dagger}\\\label{eq: density operator local cloner}
&= \int dx_i dp_i \hspace{1.5mm}G_1(x_i,p_i)
\hspace{1.5mm} \ket{\psi(x,p)} \bra{\psi(x,p)}
\end{align}
for $i = 1,2,3,4$ where $x$ and $p$ are the phase space coefficients for position and momentum respectively and $G$ is a two-dimensional Gaussian distribution given by
\begin{align}\nonumber
G(x_i,p_i) &= \frac{1}{4 \pi^2 \sqrt{V_1V_2V_3V_4}}\times\\\nonumber &{\rm exp}(-x_3^2/2V_1-p_3^2/2V_2-x_4^2/2V_3-p_4^2/2V_4)\\
&=\frac{1}{4 \pi^2} {\rm exp}(-x_3^2/2-p_3^2/2-x_4^2/2-p_4^2/2)
\end{align}
where we have used $V_1=V_2=V_3=V_4=1$ corresponding to the one unit of vacuum noise added by the local e-cloner. Eq.~(\ref{eq: density operator local cloner}) physically describes a pure EPR state $\ket{\psi}$ that is centered at the origin and has become mixed due to the one unit of noise added by the e-cloner, where this ``mixedness" is created by randomly displacing the state according to some Gaussian probability distribution with all the displacement possibilities being integrated out. In reality, our initial pure EPR state to be copied $\ket{\psi}$ is not centered at the origin but rather randomly displaced. However, if we assume the EPR state is centered at the origin, this will not change the value of the fidelity but will simplify the calculations. This is because the local e-cloner is fixed under unity gain and hence the amplitude of the two states we are comparing is the same. This means the fidelities will also not change.

An entangled state $\ket{\psi}$ centered at the origin with no displacements is given by
\begin{align}\nonumber
\ket{\psi} &= \frac{1}{\sqrt{2\pi}}\int dx_1 dx_2
e^{-x_2^{2}/4 V_S} e^{-x_1^2 V_S/4}\times\\\label{eq: EPR state fidelity}
&\ket{\frac{1}{\sqrt2} (x_2 +
x_1)}_1
\ket{\frac{1}{\sqrt2} (x_2 - x_1)}_2
\end{align}
An unknown EPR state $\ket{\psi(x,p)}$ created by displacing both arms of the original entanglement $\ket{\psi}$ can be written in the Schrodinger picture as
\begin{align}
\ket{\psi(x,p)} = \hat{D}_1(x_3,p_3) \otimes \hat{D}_2(x_4,p_4) \ket{\psi}
\end{align}
where $\hat{D}_1$ and $\hat{D}_2$ are the displacement operators used to shift both arms of the entanglement in phase space by an amount $(x_1,p_1)$ and $(x_2,p_2)$ respectively.

\begin{figure}[!ht]
\begin{center}
\includegraphics[width=8cm]{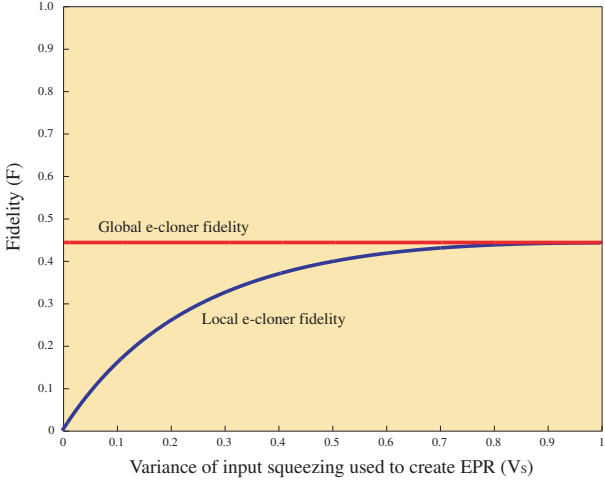}%figure_results
\caption{Plots of the fidelities for both e-cloners where both cloning machines are operating at unity gain (i.e. $g^{\pm} = \sqrt2$). The fidelities are plotted for both cloning machines: the local e-cloner fidelity $\mathcal{F_L}$ (blue line) is given by Eq.~(\ref{eq: fidelity local}) and the global cloner fidelity is $\mathcal{F_G}=4/9$ (red line). Here the bottom axes are the input squeezing used to create the entanglement with $V_S = 0$ being perfect squeezing and $V_S = 1$ being a coherent state. The global e-cloner outperforms the local e-cloner in terms of how well it copies the original entangled state.}\label{figure_results_2}
\end{center}
\end{figure}

The displacement operators are defined as $\hat{D}(x_3,p_3) = e^{i x_3 p_3/4} e^{-ix_3\hat{p}_3/2} e^{i p_3 \hat{x}_3/2}$ and $\hat{D}(x_4,p_4) = e^{i x_4 p_4/4} e^{-ix_4\hat{p}_4/2} e^{i p_4 \hat{x}_4/2}$ and act on each arm of the EPR state as follows
\begin{align}\nonumber
\hat{D}(x_3,p_3)\ket{\frac{1}{\sqrt2} (x_2 +
x_1)}_1 &= e^{ix_3 p_3/4 + i p_3 (x_2 + x_1)/2\sqrt2}\times\\
 &\ket{\frac{1}{\sqrt2}(x_2 + x_1)+ x_3}_1
\end{align}
and likewise for the other arm
\begin{align}\nonumber
\hat{D}(x_4,p_4)\ket{\frac{1}{\sqrt2} (x_2 -
x_1)}_2 &= e^{ix_4 p_4/4 + i p_4 (x_2 - x_1)/2\sqrt2}\times\\
&\ket{\frac{1}{\sqrt2}(x_2 - x_1)+ x_4}_2
\end{align}
Thus using the above definitions, our randomly displaced entangled state can be written as $\ket{\psi(x,p)}= \hat{D}(x_3,p_3,x_4,p_4)\ket{\psi}$, i.e.
\begin{align}\nonumber
\ket{\psi(x,p)} &= \frac{1}{\sqrt{2\pi}}\int dx_1 dx_2 \hspace{1.5mm}
e^{i(x_3p_3+ x_4p_4)/4 + ip_3(x_2+x_1)/(2\sqrt2)}\\\nonumber
& e^{ip_4(x_2-x_1)/(2\sqrt2)-x_2^{2}/4 V_S -x_1^2 V_S/4}\hspace{1.1mm} \times\\\nonumber
&\times \ket{\frac{1}{\sqrt2} (x_2 + x_1)+x_3}_1
\ket{\frac{1}{\sqrt2} (x_2 - x_1)+x_4}_2
\end{align}
Using the above equation (and its conjugate) along with Eq.~(\ref{eq: EPR state fidelity}), the fidelity expression of Eq.(\ref{eq: fidelity}) is calculated as
\begin{align}\nonumber
\mathcal{F_L} &= \int dx_i dp_i \hspace{1.5mm}G(x_i,p_i)
\hspace{1.5mm} |\langle \psi \ket{\psi(x,p)}|^2 \\\nonumber
&=\frac{1}{4\pi^2}\int dx_i dp_i \hspace{1.5mm} e^{-x_3^2/2-p_3^2/2-x_4^2/2-p_4^2/2} \hspace{1.5mm}\times\\\nonumber
&e^{-\frac{1}{8 V_S}\Big[(V_S^2+1) [x_3^2+ p_3^2+x_4^2 + p_4^2] -2 (V_S^2-1) [x_3 x_4- p_4 p_3]\Big]}\\\label{eq: fidelity local}
&= \frac{4 V_S}{(V_S+2)(2 V_S+1)}
\end{align}
 The fidelity of the local e-cloner is plotted in Fig.~(\ref{figure_results_2}). It starts off with a fidelity of $\mathcal{F_L}=0$ when we have infinite squeezing in the (unphysical) case of a maximally entangled CV EPR state. In this situation the fidelity is zero, because even though perfect squeezing means the EPR state is created from squeezed eigenstates (and are thus perfectly measurable), the initial random displacements mean we have no idea where in the continuous spectrum they lie, otherwise we would always have perfect fidelity (i.e. $\mathcal{F}=1$). Hence this is why we have the requirement that the EPR states are first randomly displaced in phase space. As the strength of the initial EPR state gets weaker and weaker (and finally reaches the no entanglement case of coherent states $V_S=1$) the fidelity asymptotes to the fixed fidelity of the global cloning machine which we will see is $\mathcal{F_G}=4/9$.

\subsubsection{Global E-Cloner Fidelity}

The fidelity calculation for the global e-cloner consists of comparing the entangled input state described in Fig.~\ref{figure_global_cloner}(a) as $\hat{X}_{epr1}^{\pm}$ and $\hat{X}_{epr2}^{\pm}$ to the cloned output state of $\hat{X}_{epr1A}^{\pm}$ and $\hat{X}_{epr1B}^{\pm}$. As we have seen, the global e-cloner consists of two linear cloning machines used to clone the coherent states created from un-squeezing the disentangled arms of the input state. It has been show \cite{And05} that such a linear cloning machine optimally clones a coherent state with a fidelity of $\mathcal{F}=2/3$. Now if the squeezing parameter is known, the best way to clone a squeezed state is to first un-squeeze it, clone the resulting coherent state and then squeeze again. In this case the optimal fidelity of such cloned states is also $\mathcal{F}=2/3$ \cite{Cer00}.

Lets first show that the cloning of a squeezed state with known squeezing parameter has the same fidelity of a cloned coherent state (c.f. Figure \ref{figure_global_cloner} (b)). Starting off with a pure squeezed state $\ket{\alpha_1, r}$ which goes through a squeeze gate that un-squeezes it into a coherent state given by $\ket{\alpha_1} = \hat{S}^{\dagger} \ket{\alpha_1, r}$. The density operator of this state is given by $\hat{\rho}_2$ and is then inserted into the linear cloning machine which evolves the coherent state via the CV cloning operation $\hat{C}$. The mixed output states can be described by the density operator $\hat{\rho}_3 = \hat{C} \hat{\rho}_2 \hat{C}^{\dagger}$. These cloned states are squeezed again by the same amount they were un-squeezed to give mixed squeezed states given by $\hat{\rho}_{1A} = \hat{S} \hat{\rho}_3 \hat{S}^{\dagger}$. Therefore the fidelity we are interested in is $\mathcal{F} = \bra{\alpha_1, r} \hat{\rho}_{1A} \ket{\alpha_1, r}$ and can be rewritten as
\begin{align}\nonumber
\mathcal{F} &= \langle \alpha_1, r| \hat{\rho}_{1A} \ket{\alpha_1, r}\\\nonumber
&=\langle \alpha_1, r| \hat{S} \hat{\rho}_3 \hat{S}^{\dagger} \ket{\alpha_1, r}\\
&=\langle \alpha_1| \hat{\rho}_3 \ket{\alpha_1}\\\nonumber
&= \frac{2}{3}
\end{align}
where we have used the fact that the fidelity of cloning a coherent state is $\mathcal{F} = \langle \alpha_1| \hat{\rho}_3 \ket{\alpha_1} = 2/3$.

Because we have two entangled arms to clone, the system containing both clones before the final beamsplitter, can be described as $\rho_{1A} \otimes \rho_{1B}$ where $\rho_{1B}$ is symmetrically the same as $\rho_{1A}$ and describes the other cloned squeezed state. The fidelity of such a system is then given by
\begin{align}\nonumber
\mathcal{F} &= \bra{\alpha_2, r} \bra{\alpha_1, r} \rho_{1A} \otimes \rho_{1B} \ket{\alpha_1, r} \ket{\alpha_2, r}\\
&= \frac{2}{3} \times \frac{2}{3} = \frac{4}{9}
\end{align}
where $\ket{\alpha_2, r}$ is the other input squeezed state that was originally created from disentangling the EPR state. Then putting the two mixed squeezed states through the final $50/50$ beamsplitter, to create the cloned EPR state, does not change the fidelity because the beamsplitter gate is a unitary operation. Therefore the fidelity of the global e-cloner is given by $\mathcal{F_G} = 4/9$ and is plotted in Fig.~(\ref{figure_results_2}). We can see that the fidelity of the global e-cloner asymptotes to the local e-cloner fidelity in the case of having coherent states as the input. In this situation, both e-cloners reduce to being the same quantum cloning machine.

\section{Conclusion}

In conclusion, we have considered the quantum cloning of continuous variable entangled states. We constructed and optically implemented two symmetric $1 \rightarrow 2$ entanglement cloning machines, known as e-cloners. The first, a local e-cloner, individually clones each arm of the EPR state and can be created using simple linear optical elements, such as beamsplitters and homodyne detection. The second cloning machine, known as a global e-cloner, takes the whole EPR state as an input, and then outputs two imperfect copies. It uses the same optical elements as the local e-cloner but with the requirement of three in-line squeeze gates for each clone. We considered the situation where both e-cloners copied the entangled input states with the best fidelity possible whilst determining how well the entanglement was preserved under such conditions.

We found that the local e-cloner always leads to complete destruction of the entanglement. This is to be expected as it is this feature that bounds eavesdropping in continuous variable quantum cryptography \cite{Gros03} and prevents non-causal effects in Bell measurements \cite{WZ82}. On the other hand, we find that entanglement can be (partially) preserved in the clones by the global e-cloner provided the entanglement is initially sufficiently strong. Perhaps surprisingly, we find that even the stronger EPR correlation of the entanglement can be preserved on the clones, again provided the original correlations are strong enough. From an experimental point of view, these results show that greater than 3dB of squeezing would be needed to exhibit entanglement from the global e-cloner in terms of the inseparability criterion and more than 5.7dB of squeezing for the global e-cloner in the case of the EPR paradox criterion.

Reflecting this, we find that the global e-cloner has a fixed fidelity of $\mathcal{F_G}=4/9$ whilst the local e-cloner's fidelity is a function of the input squeezing used to create the entangled states and is always less then $4/9$ and only asymptotes to the global cloner's fidelity in the case of a coherent state input. Future work might involve investigating whether the squeezed gates in the global e-cloner can be moved off-line or reduced in number and also the effect that a non-Gaussian e-cloner would have on cloning continuous variable entangled states.

We acknowledge the support of the Australian Research Council (ARC).

\end{document}